\documentstyle[emulateapj]{article}

\newcommand{\co}{\rm CO}
\newcommand{\hh}{{\rm H}_2}

\newcommand{\kps}{\,\textstyle\rm{km~s}^{-1}}
\newcommand{\lya}{Ly$\alpha$ }
\newcommand{\Mpc}{\,\textstyle\rm{Mpc}}
\newcommand{\coc}{CO(3$\rightarrow$2) }

\newcommand{\yr}{\,\textstyle\rm{yr}}

\newcommand{\msun}{\,M_{\odot}}

\newcommand{\jy}{\,\textstyle\rm{Jy}}

\newcommand{\Kkpspc}{\,\rm{K}\,\rm{km~s}^{-1}\,{\rm pc}^{2}}

\lefthead{Frayer et al.}
\righthead{Molecular Gas in SMM\,02399--0136}

\begin{document}

\title{Molecular Gas in the {\scriptsize z} =2.8 Submillimeter Galaxy
SMM~02399--0136}

\author{D. T. Frayer\altaffilmark{1}, R. J. Ivison\altaffilmark{2,3},
N. Z. Scoville\altaffilmark{1}, M. Yun\altaffilmark{4},
A. S. Evans\altaffilmark{1}, Ian Smail
\altaffilmark{5}, A. W. Blain \altaffilmark{6}, and J.--P. Kneib
\altaffilmark{7}}

\altaffiltext{1}{Astronomy Department, California Institute of Technology
105--24, Pasadena, CA  91125, USA} 

\altaffiltext{2}{Institute for Astronomy,  Department of Physics \&
Astronomy, University of Edinburgh, Blackford Hill, Edinburgh, EH9 3HJ, UK}

\altaffiltext{3}{Department of Physics \& Astronomy, University College
London, Gower Street, London, WC1E 6BT, UK}

\altaffiltext{4}{National Radio Astronomy Observatory, P.O. Box 0,
Socorro, NM  87801, USA}

\altaffiltext{5}{Department of Physics, University of Durham, South
Road, Durham, DH1 3LE, UK}

\altaffiltext{6}{Cavendish Laboratory, Madingley Road, Cambridge, CB3 OHE, UK}

\altaffiltext{7}{Observatoire Midi--Pyr\'{e}n\'{e}es, 14 Avenue E. Belin,
F--31400 Toulouse, France}

\begin{abstract}

We report the detection of \coc emission from the
submillimeter--selected hyperluminous galaxy SMM\,02399--0136.  This
galaxy is the brightest source detected in the recent SCUBA surveys of
the submillimeter sky.  The optical counterpart of the submillimeter
source has been identified as a narrow-line AGN/starburst galaxy at
$z=2.8$.  The CO emission is unresolved, $\theta \la 5\arcsec$, and is
coincident in redshift and position with the optical counterpart. The
molecular gas mass derived from the CO observations is $8\times 10^{10}
h_{75}^{-2} M_{\sun}$, after correcting for a lensing amplification
factor of 2.5.  The large CO luminosity suggests that a significant
fraction of the infrared luminosity of SMM\,02399--0136 arises from star
formation.  The high inferred star--formation rate of
$10^{3}\,M_{\sun}\yr^{-1}$ and the large gaseous reservoir may suggest
that we are seeing the formation phase of a massive galaxy.  Future CO
observations of other galaxies detected in deep submillimeter surveys
will test the generality of these conclusions for the bulk of the faint
submillimeter population.

\end{abstract}


\section{INTRODUCTION}

Recent observational results have dramatically increased our knowledge
of galaxy evolution at high redshift.  Ultraviolet-- and
optically--selected samples of distant galaxies suggest that the
star--formation rate per co--moving volume in the Universe peaks at
$z\sim 1$--2 (Madau et al.\ 1996, and references therein).  The first
results from deep surveys of the submillimeter sky (Smail, Ivison \&
Blain 1997; Blain et al. 1998; Barger et al.\ 1998; Hughes et al.\ 1998;
Eales et al. 1998) using the Submillimeter Common--User Bolometer Array
(SCUBA) on the James Clerk Maxwell Telescope may imply that a large
proportion of the star formation at high redshift is obscured by dust at
UV/optical wavelengths.  Observations of CO and dust at high redshift
are, hence, important for constraining the early evolution of galaxies.

Despite considerable observational effort, there are still only a few
known high--redshift CO sources.  These include IRAS F\,10214+4724
(Brown \& Vanden Bout 1991; Solomon, Downes, \& Radford 1992), the
Cloverleaf quasar H\,1413+117 (Barvainis et al. 1994), BR\,1202-0725
(Ohta et al. 1996; Omont et al. 1996), 53W002 (Scoville et al. 1997a),
and BRI\,1335-0417 (Guilloteau et al. 1997).  In this {\it Letter}, we
report the detection of \coc emission in submillimeter source
SMM\,02399--0136, hereafter SMM02399.

The galaxy SMM02399 is the brightest submillimeter source discovered
during a survey through rich, lensing clusters (Smail et al. 1997).
Follow--up optical work places SMM02399 at a redshift of
$z=2.803\pm0.003$, indicating that the source is gravitationally
amplified by the foreground cluster by a factor of 2.5 (Ivison et
al. 1998; hereafter I98).  The optical data show the presence of a
bright compact source (L1) containing a narrow--line AGN, with an
unknown contribution due to star formation, and a diffuse companion (L2)
thought to be interacting or merging with L1 (I98).  The submillimeter
data are consistent with thermal dust emission and suggest a massive
dust reservoir of order $10^{8}\msun$ (I98).

\section{OBSERVATIONS}

SMM02399 was observed using the Owens Valley Millimeter Array in two
configurations of six 10.4m telescopes.  A total of 54.7 hours of usable
integration time on source was obtained between January 1998 and May
1998.  The phase center for the radio observations was the position of
the brightest optical component of the SMM02399 system (L1);
RA(J2000)=$02^{\rm h} 39^{\rm m}51\fs88$; DEC(J2000)=$-01\arcdeg
35\arcmin 58\farcs0$ (I98).  The \coc line was observed using a digital
correlator configured with $112\times4$~MHz channels.  The receivers
were initially tuned to 90.927 GHz, corresponding to \coc emission at
the optical redshift of $z=2.803$.  The CO line was placed in the lower
side--band (LSB) for which the receivers were optimized.  Typical LSB
system temperatures were approximately 300 K, corrected for telescope
losses and the atmosphere.  In addition to the CO line data, we recorded
the 3mm continuum data with a 1 GHz bandwidth measured in both the lower
and upper side--bands.  The nearby quasar 0238--084 was observed every
25 minutes for gain and phase calibration.  Absolute flux calibration
was determined from observations of Uranus and Neptune, resulting in a
calibration uncertainty of about 15\% for SMM02399.

\section{RESULTS}

Figure 1a shows the \coc spectrum for the data taken from 01 January to
17 April (38.3 hr).  These data suggest that the CO emission is centered
on a slightly higher redshift ($z\simeq 2.808$) than the optical
redshift.  We obtained follow--up observations from 28 April to 21 May
(16.4 hr) with the spectrometer shifted to a central frequency of 90.815
GHz in order to obtain a complete profile for the CO emission line
(Fig. 1b).  Both data sets were combined to produce the cumulative \coc
spectrum in Figure 1c.  To produce the spectra, naturally--weighted
image cubes were made by averaging the data over independent 24 MHz (79
$\kps$) channels.  The resulting data cubes were then hanning smoothed,
and a continuum level of 0.5 mJy (i.e., a DC offset) was subtracted from
the data.  All spectra were taken along the frequency axis of the image
cubes at the position of the CO emission (Table~1).

Although the spectral--line data may suggest a weak continuum level, we
lack the sensitivity to detect continuum emission from SMM02399.  In the
LSB continuum data we find a flux density of $0.9\pm0.3$ mJy which is
consistent with only the CO emission averaged over the 1 GHz continuum
bandwidth.  From the upper--side band continuum data (line--free), we
achieve an upper limit of S(3mm) $< 0.7$ mJy ($1\sigma$).  For the
remaining of the paper, we adopt the expected continuum level of 0.3
mJy, derived from extrapolating the previous submillimeter observations
of thermal dust emission (I98).

Figure 2 shows the naturally--weighted channel maps averaged over
independent 112 MHz channels.  As stated earlier, the data were taken at
two different frequencies.  The rms noise level is higher for end
channels A\&E, since these contain less integration time than the
central three channels.  We detect CO emission at the optical position
of SMM02399 in channels C\&D.  By averaging the data in channels C\&D
(224 MHz, 740$\kps$), we achieve an 8$\sigma$ detection (Fig. 2).  We
obtain an integrated flux of $3.3 \pm0.4 \jy\kps$ and determine the CO
source position (Table~1) by fitting a Gaussian to the peak in the
integrated map.  After subtracting the expected continuum level of 0.3
mJy, the resulting \coc line flux is $S(\co) = 3.0 \pm 0.4 \jy\kps$.  By
assuming a lensing amplification factor of 2.5, the observed \coc line
flux implies intrinsic line luminosities of $L(\co) = 2.5\times 10^{7}
h_{75}^{-2} L_{\sun}$ and $L^{\prime}(\co) = 1.9\times 10^{10}
h_{75}^{-2} \Kkpspc$ ($q_o =0.5$).  $L^{\prime}(\co)$ is related to the
mass of molecular gas (including He) by $M(\hh)/L^{\prime}(\co) =
\alpha$.  We adopt a value of $\alpha =4\msun(\Kkpspc)^{-1}$, which is
similar to the Galactic value (e.g., Sanders, Scoville, \& Soifer 1991),
and infer a molecular gas mass of $M(\hh) \simeq 8 \times 10^{10}
h_{75}^{-2} M_{\sun}$.  The appropriate value for $\alpha$ is uncertain
within a factor of 2--3.  It has been argued that the Galactic value may
overestimate the amount of molecular gas in ultraluminous galaxies
(Solomon et al. 1997), while other factors such as sub--solar
metallicities (Wilson 1995) and $T_b(3-2)/T_{b}(1-0)$ brightness
temperature ratios less than unity (Devereux et al. 1994) lead to
underestimates of the molecular gas mass.

The CO emission seen in Figure~2 is unresolved.  By varying the (u,v)
weighting schemes and using deconvolution techniques, we find an
upper--limit for the CO source size of $\theta < 5 \arcsec$.  This
angular limit corresponds to a maximum linear diameter of $D < 25
h_{75}^{-1}$ kpc at $z=2.8$, or $D < 10 h_{75}^{-1}$ kpc after
correcting for lensing.  We can use this limit to constrain the
dynamical mass in SMM02399.  The dynamical mass within the CO emission
regions is $M_{dyn}= R(\Delta V/[2\sin(i)])^2/G$, where $\Delta V$ is
the observed line width of $710 \kps$.  The inclination is unknown.  We
find $M_{dyn} < 1.5 \times 10^{11} \sin^{-2}(i)\,h_{75}^{-1} M_{\sun}$.
These results indicate a gas fraction of $M(\hh)/M_{dyn} > 0.5
\sin^{2}(i)\,h_{75}^{-1}$, which is consistent with a gas--rich young
starburst.

\section{DISCUSSION}

The CO emission is positionally coincident with the SCUBA source
SMM02399 and the compact optical counterpart (L1) within the
uncertainties of the data sets (Fig. 2).  The central frequency of the
CO emission is slightly redshifted with respect to the mean redshift of
the optical lines ($400\kps$), but such offsets have been noted
previously for low redshift luminous galaxies (Mirabel \& Sanders 1988).
These offsets have been interpreted as a signature of outflows in the
optically emitting gas where the receding emission is obscured by dust
(Mirabel \& Sanders 1988).  The CO emission is unresolved ($\theta <5
\arcsec$), unlike the \lya emission and the 20 cm radio continuum
emission which are $\theta \sim 8 \arcsec$ in extent (I98).  Since it is
common for ultraluminous merger systems to be compact CO sources (Downes
\& Solomon 1998), it is not surprising that the CO emission for SMM02399
is unresolved.  In addition, the broad CO line, with an apparent double
peak profile, is consistent with the suspected merger scenario for the
system (I98).

The derived molecular gas mass of approximately $10^{11} M_{\sun}$ for
SMM02399 is about 50 times higher than the Milky Way (Solomon \& Rivolo
1989) and is a few times more massive than the most luminous low
redshift infrared galaxies (Sanders et al. 1991).  The high molecular
gas mass observed in SMM02399 is not unique for high--redshift systems.
Other high--redshift CO sources have similarly large molecular gas
masses (e.g., Guilloteau et al. 1997), and this may indicate evolution
in $M(\hh)$.  Such increases are expected in simple theoretical models
(Frayer \& Brown 1997).  However, given the uncertainties in $\alpha$
(Solomon et al. 1997), the lensing uncertainties, and the small number
of high--redshift CO sources, the possibility of evolution is only a
suggestion at the present time.

We derive a gas to dust ratio of $M(\hh)/M({\rm dust}) \simeq 140$--700
for SMM02399, assuming a range of dust temperatures of 30--70~K.  This
gas to dust ratio is similar that found for other high--redshift CO
sources and is also similar to that observed in nearby galaxies
(Devereux \& Young 1990).  This consistency suggests that the
high--redshift CO sources are already significantly chemically evolved.
Based on the deficient CO luminosities observed in metal--poor galaxies
(e.g., Sage et al. 1992), the $L^{\prime}(CO)/L(1.25{\rm mm})$
luminosity ratios for the high--redshift CO sources are consistent with
metallicities of $Z\ga 0.2 Z_{\sun}$.
 
The SCUBA measurements indicate that SMM02399 is a hyperluminous
infrared galaxy (HyLIG, as defined by Sanders \& Mirabel 1996), and the
optical data suggest the presence of both a dust--enshrouded AGN and
starburst activity (I98).  An important unanswered question is which
mechanism is responsible for the majority of its immense infrared
luminosity.  The relative importance of AGN and starburst activity seen
in even the nearest ultraluminous galaxies is still uncertain (e.g.,
Scoville, Yun, \& Bryant 1997b; Downes \& Solomon 1998).  Based on
observations using the Infrared Space Observatory, Genzel et al. (1998)
find that about 25\% of ultraluminous galaxies are powered by AGNs,
while the remaining 75\% are dominated by star formation.  It is not
clear if this trend will hold for the HyLIGs.

The detection of CO in SMM02399 suggests the importance of star
formation by showing the presence of a massive reservoir of molecular
gas from which stars are formed.  Other HyLIG galaxies which do not show
CO have extremely high $L({\rm IR})/L^{\prime}(\co)$ ratios and are
thought to be dominated by AGN activity (Evans et al. 1998).  The
$L({\rm IR})/L^{\prime}(\co)$ ratio for SMM02399, on the other hand, is
roughly consistent with that found for luminous starbursts (Sanders \&
Mirabel 1996).  After correcting for the lensing of SMM02399, we find a
far--infrared (FIR) luminosity (Helou et al. 1988) of $L({\rm FIR}) =
1.2\times 10^{13} h_{75}^{-2}\,L_{\sun}$ and a ratio of $L({\rm
FIR})/L^{\prime}(\co)= 630 \,L_{\sun} (\Kkpspc)^{-1}$.  For comparison,
the ultraluminous putative starburst galaxy Arp 220 has a $L({\rm
FIR})/L^{\prime}(\co)$ ratio which is only about a factor of 2 lower.

The starburst nature of SMM02399 is further supported by the
far--infrared to 4.85 GHz radio flux ratio parameter $q$ (Condon,
Frayer, \& Broderick 1991)\footnote{We use a $q$ parameter based on the
4.85 GHz radio flux densities instead of the more often cited $q$
parameter based on a radio frequency 1.4 GHz, since the radio
rest--frame frequency observed for SMM02399 is 5.3 GHz (I98).}.  For
SMM02399 we find $q=2.3$ which is similar to, but lower than, the mean
value of $\langle q\rangle=2.64\pm0.16$ observed in nearby starburst
galaxies (Condon et al. 1991).  Although the data for SMM02399 show the
importance of star formation, the data also suggest that the AGN
activity is not negligible.  For pure star formation, we would expect a
higher $q$ value and a lower $L({\rm FIR})/L^{\prime}(\co)$ ratio.  In
summary, the current data are consistent with $50\pm25$\% of the
infrared luminosity being powered by star formation, with the remaining
fraction due to the AGN.

Assuming that half of the infrared luminosity of SMM02399 is powered by
star formation, the implied star formation rate for massive stars is
SFR$(M>5\msun) \simeq 550 \msun \yr^{-1} h_{75}^{-2}$, using the
relationship given by Condon (1992).  Including low mass stars, we
expect a total SFR of order $10^{3}\msun \yr^{-1}$, depending on the
details of the IMF.  Although high, this SFR is below the maximum
allowable rate permitted for the observed CO line width (Lehnert \&
Heckman 1996).

\section{CONCLUSIONS}

We report the first detection of CO in a distant galaxy, SMM\,02399,
discovered during the recent deep surveys of the submillimeter sky.  The
CO emission is coincident in position and redshift with the optical
counterpart of SMM02399.  The CO observations indicate the presence of
$8\times 10^{10} h^{-2}_{75} M_{\sun}$ of molecular gas.  Comparison of
the properties of SMM02399 with local starbursts suggests that star
formation accounts for approximately 25--75\% of its immense infrared
luminosity, with remaining contribution coming from an AGN.  The derived
star--formation rate is of order $10^{3}\msun\yr^{-1}$.  Similar
observations of other high--redshift submillimeter sources will test
whether the properties of SMM02399 are representative of the general
faint submillimeter population, as well as placing limits on the
contribution of dust--enshrouded AGN to the extragalactic submillimeter
background.

\acknowledgments

We thank our colleagues at the Owens Valley Millimeter Array who have
helped make these observations possible.  The Owens Valley Millimeter
Array is a radio telescope facility operated by the California Institute
of Technology and is supported by NSF grants AST 93--14079 and AST
96--13717.  RJI and IRS acknowledge support from PPARC Advanced
Fellowships.

\newpage

\begin{deluxetable}{ll}
\tablecaption{CO Observational Results}
\tablewidth{300pt}  
\tablehead{\colhead{Parameter}&\colhead{Value}}
\startdata
RA (J2000)  & $02^{\rm h} 39^{\rm m}51\fs89\pm0\fs02$ \nl
DEC (J2000) &$-01\arcdeg 35\arcmin 58\farcs9\pm0\farcs5$ \nl
$\langle z(\co) \rangle$  &$2.808\pm0.002$ \nl
$\Delta V(\co)_{FWHM}$& $710\pm80\kps$\nl
$S(\co)$\tablenotemark{a}& $3.0\pm0.4 \jy\kps$ \nl
$L(\co)$\tablenotemark{b}&  $2.5\times 10^{7} h_{75}^{-2}
L_{\sun}$ \nl 
$L^{\prime}(\co)$\tablenotemark{b}& $1.9\times 10^{10}
h_{75}^{-2}\Kkpspc$\nl     
$M(\hh)$\tablenotemark{b,c} & $8\times10^{10} h_{75}^{-2} M_{\sun}$\nl
\enddata 
\tablenotetext{a}{Observed \coc line flux assuming a continuum level of
0.3 mJy.}
\tablenotetext{b}{Intrinsic value assuming a lensing amplification
factor of 2.5, $q_o =0.5$, and H$_o = 75\,h_{75} \kps \Mpc^{-1}$.}
\tablenotetext{c}{Estimated using $\alpha =4 \msun (\Kkpspc)^{-1}$.}
\end{deluxetable}

\newpage

\begin{figure}
\includegraphics{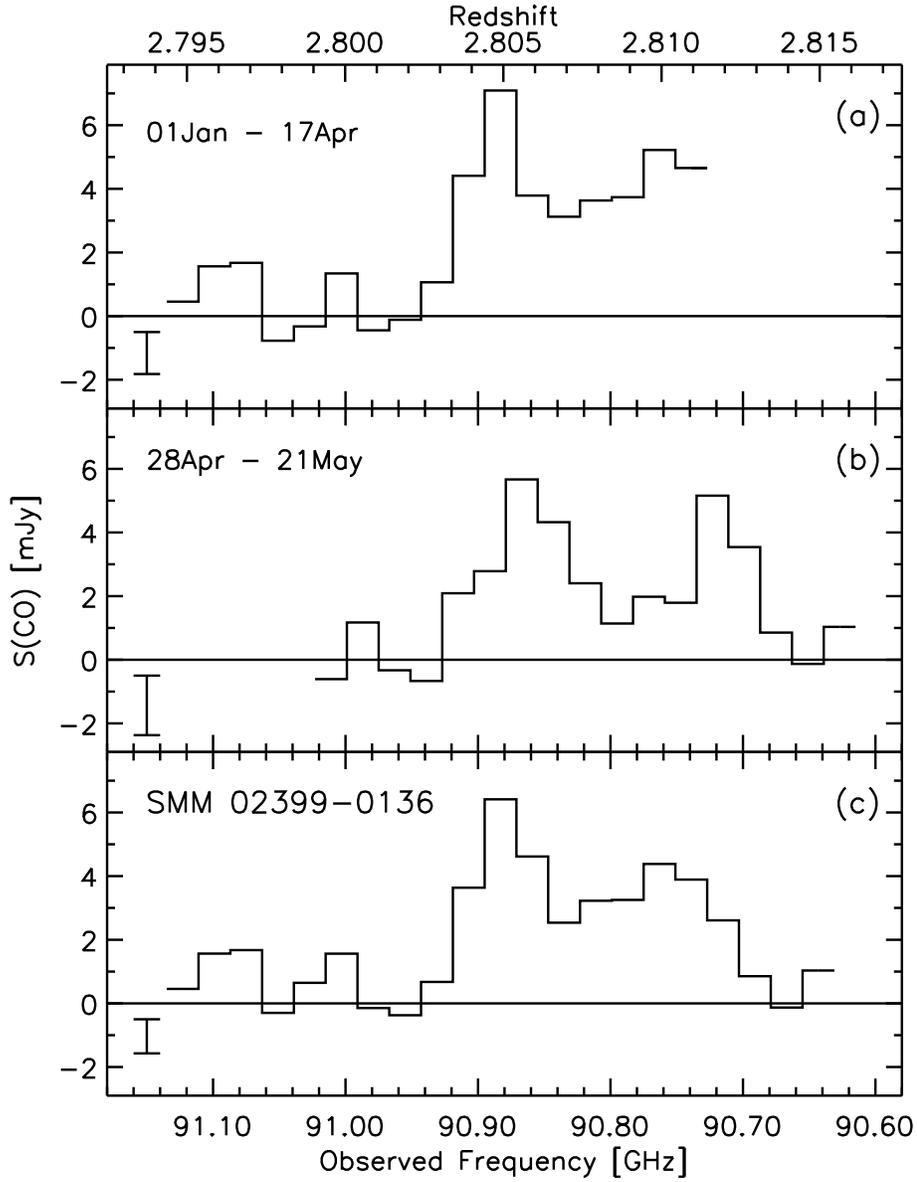} \vspace*{7.0in}
\caption{The \coc spectra for SMM02399 observed with the OVRO
Millimeter Array.  Panels (a)\&(b) show subsets of the data, while
panel (c) displays the cumulative spectrum.  The $1\sigma$ error bar is
shown in the lower left of each panel.}
\end{figure}

\newpage

\begin{figure}
\includegraphics{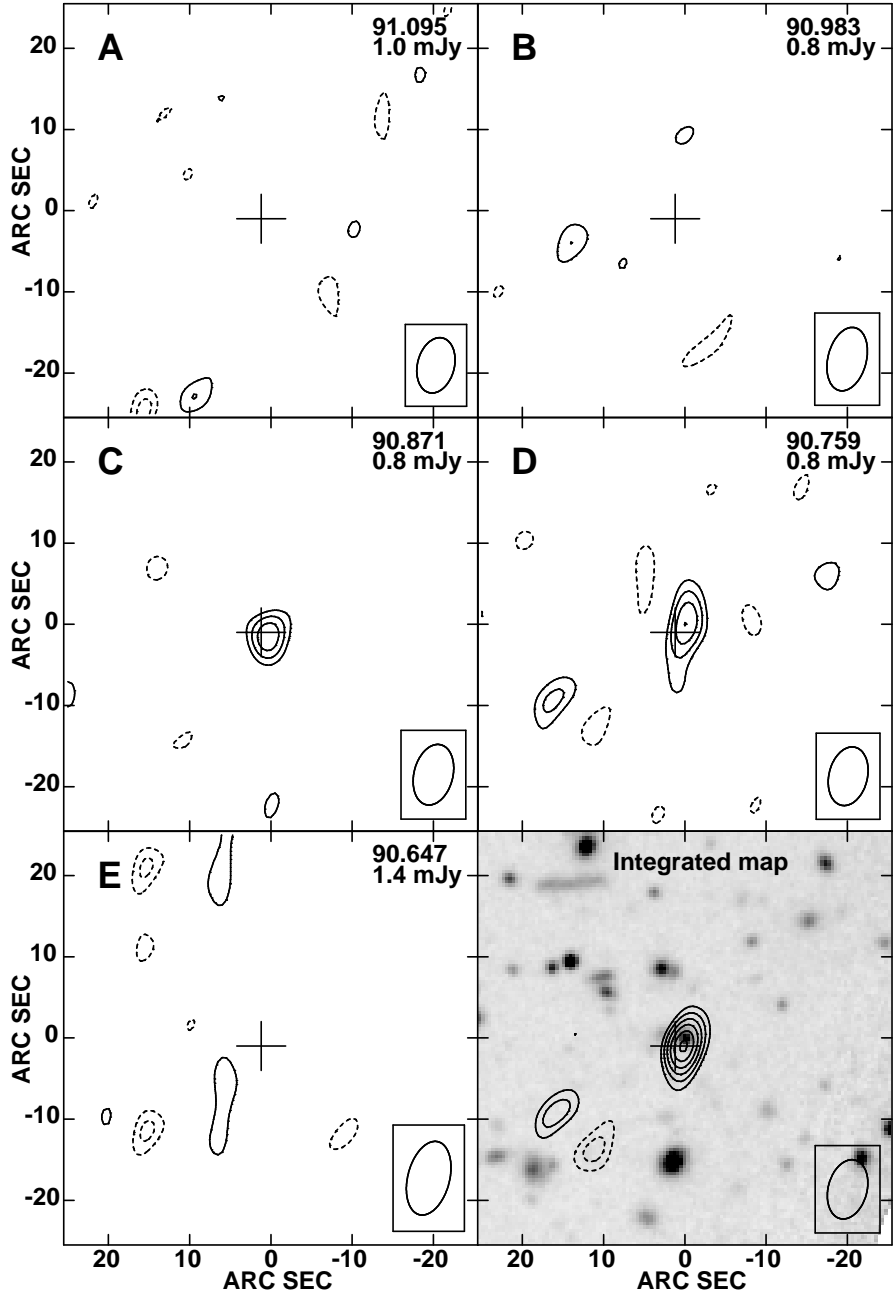} \vspace*{7.in}
\caption{In panels A--E we show the \coc channel maps averaged
over 112 MHz.  The observed central frequency in GHz and the $1\sigma$
rms level of each channel are given in the top right.  The positional
offsets are relative to the optical position (north is at the top, and
east is at the left), and the SCUBA position (I98) is shown by a cross.
The contour levels are $1\sigma\times(-3.5,-2.5,2.5,3.5,4.5,5.5)$ in
panels A--E.  In the lower right panel is the integrated CO map averaged
over 224 MHz (channels C\&D) overlaid on an optical B--band image taken
with the Canada--France--Hawaii Telescope (Kneib et al. 1994).  The
$1\sigma$ rms level is $0.4\jy\kps$, and the contour levels are
$1\sigma\times(-4,-3,3,4,5,6,7,8)$.  The upper limit to the 3mm
continuum level is $0.5 \jy\kps$ in the integrated map.  The synthesized
beam size is shown in the lower right ($7\farcs4\times4\farcs8, {\rm
PA}=-15\arcdeg$).}

\end{figure}

\end{document}